\documentclass[a4paper,aps,preprintnumbers,showpacs,amsmath,amssymb,prl]{revtex4}
\usepackage{mathrsfs}
\usepackage{amsmath}
\usepackage[dvips]{graphicx}
\usepackage{epsfig}
\usepackage[dvips]{graphics,graphicx}
\begin{document}

\title{Dynamics of a movable micro-mirror in a nonlinear optical cavity}

\author{Tarun Kumar$^{1}$, Aranya B.\ Bhattacherjee$^{2}$ and ManMohan$^{1}$}

\address{$^{1}$Department of Physics and Astrophysics, University of Delhi, Delhi-110007, India}\address{$^{2}$Department of Physics, ARSD College, University of Delhi (South Campus), New Delhi-110021, India}

\begin{abstract}
We consider the dynamics of a movable mirror (cantilever) of a nonlinear optical cavity. We show that a $\chi^{(3)}$ medium with a strong Kerr nonlinearity placed inside a cavity inhibits the normal mode splitting (NMS) due to the photon blockade mechanism. This study demonstrates that NMS could be used as a tool to observe the photon blockade effect. We also found that the backaction cooling of the movable mirror is reduced in the presence of the Kerr medium.
\end{abstract}


\maketitle

\section{Introduction}

The interaction between a movable mirror and the radiation field of an optical cavity has recently been the subject of extensive theoretical and experimental investigations. These optomechanical systems couple the mechanical motion to an optical field directly via radiation pressure buildup in a cavity. The coupling of mechanical and optical degrees of freedom via radiation pressure has been a subject of early research in the context of laser cooling \cite{hansch,wineland, chu} and gravitational-wave detectors \cite{caves}. Recently there has been a great surge of interest in the application of radiation forces to manipulate the center-of-mass motion of mechanical oscillators covering a huge range of scales from macroscopic mirrors in the Laser Interferometer Gravitational Wave Observatory (LIGO) project \cite{corbitt1, corbitt2} to nano-mechanical cantilevers\cite{hohberger, gigan, arcizet, kleckner, favero, regal}, vibrating microtoroids\cite{carmon, schliesser} membranes\cite{thompson} and Bose-Einstein condensates \cite{brennecke, murch}. The quantum optical properties of a mirror coupled via radiation pressure to a cavity field show interesting similarities to an intracavity Kerr-like interaction \cite{fabre}. Recently, in the context of classical investigations of nonlinear regimes, the dynamical instability of a driven cavity having a movable mirror has been investigated \cite{marquardt}. Theoretical work has proposed to use the radiation-pressure coupling for quantum non-demolition measurements of the light field \cite{braginsky}.

It has been shown that ground state cooling of micro-mechanical mirror is possible only in the resolved side band regime (RSB) where the mechanical resonance frequency exceeds the bandwidth of the driving resonator \cite{marquardt,braginsky}. The cooling of mechanical oscillators in the RSB regime at high driving power can entail the appearance of normal mode splitting (NMS) \cite{dobrindt}. Recently, it was shown that an optical parametric amplifier inside a cavity considerably improves the cooling of a micro-mechanical mirror by radiation pressure \cite{huang}.

In this paper, we consider the dynamics of a movable mirror interacting with a nonlinear optical cavity mode and predict novel properties of the dynamics of the system. Giant optical Kerr nonlinearities are obtained by placing a $\chi^{(3)}$ medium inside a cavity\cite{Imamoglu}. This gives rise to a strong nonlinear interactions between photons. A single photon in a cavity can block the injection of a second photon due to a photon blockade effect. We show that due to the photon blockade mechanism, as the Kerr nonlinearity is increased, the NMS progressively decreases.

\section{The Model}
We consider an optical Kerr medium with $\chi^{(3)}$ nonlinearity inside a Fabry-Perot cavity with one fixed partially transmitting mirror and one movable totally reflecting mirror in contact with a thermal bath in equilibrium at temperature $T$, as shown in Fig.1. The movable mirror is treated as a quantum mechanical harmonic oscillator with effective mass $m$, frequency $\Omega_{m}$ and energy decay rate $\Gamma_{m}$. 
The system is also coherently driven by a laser field with frequency $\omega_{L}$ through the cavity mirror with amplitude $\epsilon$. It is well known that high-Q optical cavities can significantly isolate the system from its environment, thus strongly reducing decoherence and ensuring that the light field remains quantum-mechanical for the duration of the experiment. We also assume that the induced resonance frequency shift of the cavity and the nonlinear interaction coefficient $\eta$ are much smaller than the longitudinal mode spacing, so that we restrict the model to a single longitudinal mode $\omega_{c}$. We also assume that $\Omega_{m}<<\pi c/L$ (adiabatic limit); $c$ is the speed of light in vacuum and $L$  the cavity length in the absence of the cavity field. The total Hamiltonian of the system in a frame rotating at the laser frequency $\omega_{L}$ can be written as

\begin{figure}[t]
\hspace{1.0cm}
\includegraphics [scale=0.5] {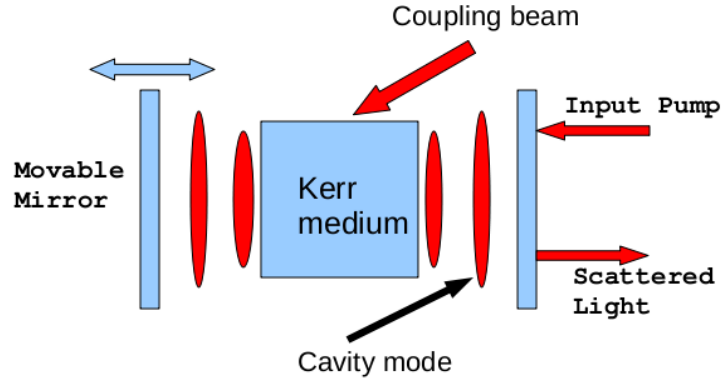} 
\caption{Optomechanical realization of parametric coupling of a mechanical oscillator to a optical mode of a nonlinear cavity.}
\label{figure1}
\end{figure}

\begin{eqnarray}\label{eff}
H&=& \hbar (\omega_{c}-\omega_{L})a^{\dagger}a-\hbar g_{m} a^{\dagger}a q+\left(\dfrac{p^2}{2m}+m\Omega_{m}^{2}q^{2} \right) \nonumber \\&+& i \hbar \epsilon (a^{\dagger}-a)+\hbar \eta a^{\dagger 2} a^{2}\,
\end{eqnarray}

Here $a$ and $a^{\dagger}$ are the annihilation and creation operators for the cavity field respectively. Also $q$ and $p$ are the position and momentum operators for the movable mirror. The parameter $g_{m}= \omega_{c}/L$ is the coupling parameter between the cavity field and the movable mirror and $\eta$ is the anharmonicity parameter and is proportional to the third-order nonlinear susceptibility $\chi^{(3)}$ of the Kerr medium: $\eta=3\hbar\omega_{c}^{2} Re[\chi^{(3)}]/2\epsilon_{0}V_{c}$, $\epsilon_{0}$ is the dielectric constant of the medium and $V_{c}$ is the volume of the cavity.  The input laser field populates the intracavity mode which couples to the movable mirror through the radiation pressure.  The field in turn is modified by the back-action of the cantilever. It is important to notice the nonlinearity in Eqn. (\ref{eff}) arising from the coupling between the intracavity intensity and the position operator of the mirror. The system we are considering is intrinsically open as the cavity field is damped by the photon-leakage through the massive coupling mirror and the mirror is connected to a bath at finite temperature. In the absence of the radiation-pressure coupling, the cantilever would undergo a pure Brownian motion driven by its contact with the thermal environment. The motion of the system can be described by the following quantum Langevin equations:

\begin{equation}
 \dot q= \dfrac{p}{m}
\end{equation}

\begin{equation}
 \dot p=-m \Omega_{m}^{2}q+\hbar g_{m} a^{\dagger} a-\Gamma_{m} p+\xi
\end{equation}

\begin{equation}
 \dot a=i(\omega_{L}-\omega_{c})a+i g_{m} q a+\epsilon -2i \eta a^{\dagger}a^{2}-\kappa a+\sqrt{2\kappa} a_{in}.
\end{equation}

Here $a_{in}$ is the input vacuum noise operator and it obeys the following correlation functions:

\begin{equation}
 <\delta a_{in}(t) \delta a^{\dagger}_{in}(t')>=\delta(t-t')
\end{equation}

\begin{equation}
 <\delta a_{in}(t) \delta a_{in}(t')>=<\delta a^{\dagger}_{in}(t) \delta a^{\dagger}_{in}(t')>=0.
\end{equation}

The force $\xi$ is the Brownian noise operator resulting from the coupling of the movable mirror to the thermal bath, whose mean value is zero, and has the following correlation function at temperature $T$:

\begin{equation}
 <\xi(t) \xi(t')>=\dfrac{\hbar \Gamma_{m} m}{2 \pi}\int \omega e^{-i \omega (t-t')}\left[\coth{\left(\dfrac{\hbar \omega}{2 k_{B} T} \right)}+1  \right] d\omega, 
\end{equation}

where $k_{B}$ is the Boltzmann constant and $T$ is the thermal bath temperature. The steady state values of $p$, $q$ and $a$ are obtained as:

\begin{equation}
 p_{s}=0,
\end{equation}

\begin{equation}
 q_{s}=\dfrac{\hbar g_{m} |a_{s}|^{2}}{m \Omega_{m}^{2}},
\end{equation}

\begin{equation}
 a_{s}=\dfrac{\epsilon}{\kappa+i(\Delta+2 \eta |a_{s}|^{2})},
\end{equation}

where $\Delta=\omega_{c}-\omega_{L}-g_{m}q_{s}$ is the effective cavity detuning which includes the radiation pressure effects. Here $q_{s}$ denotes the new equilibrium position of the mirror while $a_{s}$ denotes the steady state amplitude of the cavity field. Both $q_{s}$ and $a_{s}$ displays multistable behaviour due to the nonlinear interaction between the mirror and the cavity field.  From the above equations we clearly see how the mirror dynamics affects the steady state of the intracavity field. The coupling to the mirror shifts the cavity resonance frequency and changes the field inside the cavity in a way to induce a new stationary intensity. The change occurs after a transient time depending on the response of the cavity and strength of the coupling to the mirror.

\section{Dynamics of small fluctuations}

Here we show that the coupling of the mechanical oscillator and the cavity field fluctuations in the presence of the Kerr medium leads to the inhibition of the normal mode splitting (NMS). The optomechanical NMS however involves driving two parametrically coupled nondegenerate modes out of equilibrium. The NMS does not appear in the steady state spectra but rather manifests itself in the fluctuation spectra of the mirror displacement. To this end, we write each canonical operator of the system as a sum of its steady-state mean value and a small fluctuation with zero mean value, $a \rightarrow a_{s}+ \delta a$, $p \rightarrow p_{s}+\delta p$, $q \rightarrow q_{s}+\delta q$ and linearize to obtain the following Heisenberg-Langevin equations for the fluctuation operators

\begin{equation}\label{fluc1}
 \dot \delta q=\dfrac{\delta p}{m},
\end{equation}

\begin{equation}\label{fluc2}
 \dot \delta p= -m \Omega_{m}^{2} \delta q+\hbar g_{m} (a_{s} \delta a^{\dagger}+ a_{s}^{*} \delta a)-\Gamma_{m} \delta p+\xi,
\end{equation}

\begin{equation}\label{fluc3}
 \dot \delta a=-i \Delta \delta a +i g_{m} a_{s} \delta q-2i\eta (2 |a_{s}|^{2} \delta a-a_{s}^{2} \delta a^{\dagger})-\kappa \delta a + \sqrt{2\kappa} \delta a_{in}.
\end{equation}

Here we will always assume $\Gamma_{m}<<\kappa$. Eqns. (\ref{fluc1}, \ref{fluc2}, \ref{fluc3}) and their Hermitian conjugates constitute a system of four first order coupled operator equations, for which the Routh-Hurwitz criterion implies that the system is stable for the following conditions:

\begin{equation}
 (\Delta+4 \eta |a_{s}|^{2})^{2}+\kappa^{2}+ \Omega_{m}^{2}+2 \Gamma_{m} \kappa +2i \eta (a_{s}^{2}-a_{s}^{*2})(\kappa+\Gamma_{m})-2 \eta^{2}(a_{s}^{4}+a_{s}^{*4})>0,
\end{equation}

\begin{equation}
 \Gamma_{m}(\Delta+4 \eta |a_{s}|^{2})^{2}+2 \Omega_{m}^{2}\left\lbrace \kappa+i \eta (a_{s}^{2}-a_{s}^{*2})\right\rbrace - \Gamma_{m} \eta^{2}(a_{s}^{2}+a_{s}^{*2})^{2}>0,
\end{equation}

\begin{equation}
\Omega_{m}^{2}\left\lbrace [\kappa+i \eta (a_{s}^{2}-a_{s}^{*2})]^{2}+[\Delta+4 \eta |a_{s}|^{2}]^{2} \right\rbrace + \dfrac{\hbar g_{m}^{2} \eta}{m}(a_{s}^{2}+a_{s}^{*2})^{2}-\Omega_{m}^{2} \eta^{2}(a_{s}^{2}+a_{s}^{*2})^{2}-\dfrac{2 \hbar g_{m}^{2} |a_{s}|^{2}}{m}(\Delta+4 \eta |a_{s}|^{2})>0.
\end{equation}

The study of these conditions reveals the point at which the system enters an unstable regime. Here, we will restrict ourselves to the stable regime.

We now transform to the quadratures: $\delta x=\delta a^{\dagger}+\delta a$, $\delta y=i(\delta a^{\dagger}-\delta a)$, $\delta x_{in}=\delta a_{in}^{\dagger}+\delta a_{in}$ and $\delta y_{in}=i(\delta a_{in}^{\dagger}-\delta a_{in})$. The position fluctuations of the movable mirror in Fourier space is given by

\begin{equation}
\delta q(\omega)=\dfrac{1}{d(\omega)}\left\lbrace [(\kappa-i \omega)^{2}+\delta '^{2}] \xi (\omega)-i \hbar g_{m} \sqrt{2 \kappa}[(\omega+i \kappa+\delta)a_{s}^{*} \delta a_{in}+(\omega+i \kappa-\delta)a_{s} \delta a_{in}^{\dagger}]\right\rbrace, 
\end{equation}

where $d(\omega)=m[\Omega_{m}^{2}-\omega^{2}-i \omega \Gamma_{m}][(\kappa-i \omega)^{2}+\delta'^{2}]-2 \hbar g'^{2}_{m} \delta''$, $\delta'^{2}=\Delta'^{2}-4 \eta'^{2}$, $\Delta'=\Delta+4 \eta'$, $\eta'=\eta |a_{s}|^{2}$, $\delta=\Delta+2 \eta'$, $\delta''=\Delta'-\eta'$ and $g'_{m}=g_{m}|a_{s}|$. In the above equation for $\delta q$, the term proportional to $\xi(\omega)$ arises from thermal noise, while the term proportional to $g_{m}$ originates from radiation pressure. The displacement spectrum is obtained from

\begin{equation}
S_{q}(\omega)=\dfrac{1}{4 \pi} \int d\Omega e^{-i(\omega+\Omega)t} <\delta q(\omega) \delta q(\Omega)+\delta q(\Omega)\delta q(\omega))>,
\end{equation}

together with the correlation functions:

\begin{equation}
 <\delta a_{in}(\omega) \delta a_{in}^{\dagger}(\Omega)>=2 \pi \delta(\omega+\Omega)
\end{equation}

\begin{equation}
<\xi(\omega) \xi(\Omega)> =2 \pi \hbar \Gamma_{m} m \omega \left[ 1+coth(\dfrac{\hbar \omega}{2 k_{B}T})\right] \delta(\omega+\Omega). 
\end{equation}

The displacement spectrum in Fourier space is finally obtained as:

\begin{equation}
S_{q}(\omega)=\hbar |\chi|^{2}\left\lbrace m \Gamma_{m} \omega \coth{\left( \dfrac{\hbar \omega}{2 k_{B} T}\right) }+\dfrac{4 \kappa g'^{2}_{m}(\omega^{2}+\kappa^{2}+\delta^{2})}{(\kappa^{2}+\delta'^{2}-\omega^{2})^{2}+4 \kappa^{2} \omega^{2}}\right\rbrace, 
\end{equation}

where, 

\begin{equation}
 \chi^{-1}(\omega)=m [\Omega_{eff}^{2}-\omega^{2}]-i \omega \Gamma_{eff},
\end{equation}

\begin{equation}
\Omega_{eff}^{2}=\Omega_{m}^{2}-\dfrac{4 \hbar g'^{2}_{m} \delta'' (\kappa^{2}+\delta'^{2}-\omega^{2})}{(\kappa^{2}-\omega^{2}+\delta'^{2})^{2}+4 \kappa^{2} \omega^{2}},
\end{equation}

\begin{equation}
\Gamma_{eff}=m \Gamma_{m}+\dfrac{4 \hbar g'^{2}_{m} \delta'' \kappa}{(\kappa^{2}-\omega^{2}+\delta'^{2})^{2}+4 \kappa^{2} \omega^{2}}.
\end{equation}

\begin{figure}[t]
\hspace{1.0cm}
\includegraphics [scale=1.0] {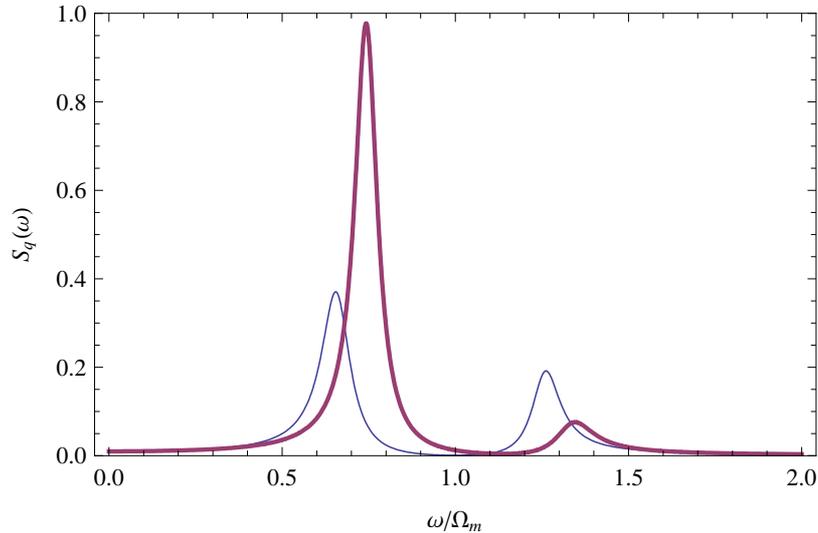} 
\caption{Plot of the displacement spectrum $S_{q}(\omega)$ for two values of the nonlinear coefficient: $\eta'=0$(thin line) and $\eta'=0.04$ (thick line). Parameters used are: $\Gamma/\Omega_{m}=0.01$, $\kappa/\Omega_{m}=0.1$, $g_{m}'/\Omega_{m}=0.1$ and $\Delta/\Omega_{m}=1$. Clearly for a finite value of $\eta$ the NMS slowly becomes less prominent.}
\label{figure2}
\end{figure}

Any information about the mirror's modified motion can be obtained from the study of $S_{q}(\omega)$. An immediate observation reveals that $S_{q}(\omega)$ is peaked at a frequency $\Omega_{eff}$. In the expression for $S_{q}(\omega)$, the first term represents the contribution due to the thermal contact with the bath while the second term is attributed to the contribution due to radiation pressure. In the absence of radiation pressure any dependence from the effective detuning $\Delta$ vanishes and the resulting spectrum is simply that of a harmonic oscillator undergoing Brownian motion at temperature $T$. In fig.2, we show the plot of $S_{q}(\omega)$ as a function of $\omega/\Omega_{m}$ for $\eta'=0$(thin line) and $\eta'=0.04$(thick line). In the absence of Kerr nonlinearity, a clear NMS is observed in the displacement spectrum. The NMS is associated to a mixing between the mechanical mode and the fluctuation around the steady state of the cavity field. In the presence of finite Kerr nonlinearity, we notice the absence of NMS. The absence of NMS in the presence of Kerr nonlinearity is understood as follows: If $\eta>>\kappa,\epsilon$, the applied field will couple the vacuum state to the Fock state with single photon resonantly. The higher lying photon-number states may be neglected since they are out of resonance. Now if initially a photon from the driving field is injected in the cavity with a probability determined by the drive strength. However, injection of a second photon will be blocked, since the presence of two photons in the cavity will require an aditional $\hbar \eta$ energy, which cannot be provided by the pump laser. Only after the first photon leaves the cavity can a second photon be injected. The strong interactions between the photons therefore causes a photon (Kerr) blockade of cavity transmission and this drastically reduces the photon number fluctuation. In any case if $\eta<\kappa,\epsilon$, the cavity would contain more than one photon and the above argument is still valid due to the photon-photon repulsion. We now return to the linearized Heisenberg-Langevin equations (\ref{fluc1}, \ref{fluc2}, \ref{fluc3}) and calculate the corresponding eigenfrequencies that determine the dynamics of NMS. In particular, we focus on the following: (i) $\kappa< \Omega_{m}/2$, (ii) $g_{m}<\Omega_{m}/2$, (iii) $\Gamma_{m}<<\kappa$ and (iv) $\Delta-\Omega_{m}<<\Omega_{m}$. The two eigenfrequencies are found to be:

\begin{equation}
\omega_{\pm}=\dfrac{\Delta_{\eta}+\Omega_{m}-i \kappa-i \Gamma_{m}}{2} \pm \sqrt{g_{m,eff}^{2}-\dfrac{\left( i(\Delta_{\eta}-\Omega_{m})+(\kappa-\Gamma_{m})\right)^{2} }{4}},
\end{equation}

where, $\Delta_{\eta}=\Omega_{m}+6 \eta'$ and $g_{m,eff}=2 g_{m} a_{s} \sqrt{\dfrac{\hbar}{m \Omega_{m}}}$. There is another pair of eigenfrequencies $-\omega_{\pm}^{*}$. For $\eta'=0$, $\Delta=\Omega_{m}$ and $\kappa>>\Gamma_{m}$, the square root term of $\omega_{\pm}$ is real for $g_{m,eff}>\kappa/2$ and shows NMS. On the other hand, for $\eta' \neq 0$, NMS is exhibited for $\Delta=\Omega_{m}-6 \eta'$.

Next we analyze the influence of the Kerr nonlinearity on the backaction cooling of the movable mirror. The effective temperature is defined by the total energy of the movable mirror, $k_{B} T=\dfrac{1}{2}m \Omega_{m}^{2}<q^{2}>+\dfrac{<p^{2}>}{2m}$ \cite{huang}, where $<q^{2}>=\dfrac{1}{2\pi}\int S_{q}(\omega)d\omega$, $<p^{2}=\dfrac{1}{2 \pi} \int S_{p}(\omega) d \omega>$ and $S_{p}(\omega)=m^{2} \omega^{2} S_{q}(\omega)$. This basically means that the effective temperature is proportional to the displacement spectrum. From Fig.2, we observe that the displacement spectrum for $\eta \neq 0$ is always more than that for $\eta=0$. From this we conclude that $T_{eff}(\eta \neq 0)$ $>$ $T_{eff}(\eta=0)$. We can come to this conclusion also from the fact that there is a reduction in the number of photons (hence radiation pressure) in the cavity due to photon-photon repulsion due to the presence of the Kerr medium and hence an increase in the temperature of the movable mirror.
An important point to note is that in order to observe the NMS, the energy exchange between the two modes(mechanical and photon number fluctuation) should take place on a time scale faster than the decoherence of each mode. Also the parameter regime in which NMS may appear implies cooling. On the negative detuning side, the observation of NMS is prevented by the onset of parametric instability.

To demonstrate that the dynamics investigated here are within experimental reach, we discuss the experimental parameters from \cite{schliesser2}:  From \cite{schliesser2},the mechanical frequency $\Omega_{m}=2 \pi \times 73.5 Mhz$ and $\Gamma_{m}=2 \pi \times 1.3 Khz$. The coupling rate $g_{m}=2 \pi \times 2.0 Mhz$. From \cite{Imamoglu}, the Kerr nonlinearity is numerically estimated to be about $\eta=100Mhz$ for extremely strong photon-photon repulsion. Here we take $\eta=2 \pi \times 3 Mhz$.
The energy of the cavity mode decreases due to the photon loss through the cavity mirrors, which leads to a reduced atom-field coupling. Photon loss can be minimized by using high-Q cavities. Our proposed detection scheme relies crucially on the fact that coherent dynamics dominate over the losses. It is important that the characteristic time-scales of coherent dynamics are significantly faster than those associated with losses (the decay rate of state-of-art optical cavities is typically 17 kHz \cite{Klinner06}).

\section{Conclusions}
In summary we have analyzed the influence of a Kerr medium on the dynamics of a micro-mechanical movable mirror. We have shown that as the Kerr nonlinearity increases, the normal mode splitting (NMS) progressively weakens. This is attributed to the photon blockade mechanism which decreases the photon fluctuations due to photon-photon repulsion. Further we found that the temperature of the micro-mechanical mirror is enhanced due to the presence of the Kerr medium.The present scheme could also be used to detect the photonic repulsion effect.

\end{document}